\documentstyle[preprint,aps,epsfig]{revtex}

\def\be{\begin{equation}}
\def\ee{\end{equation}}

\begin{document}
\draft

\title{Effective conductivity of composites of graded spherical particles}

\date{\today}
\author{K. W. Yu$^{1,2}$\footnote{Corresponding author.
 Electronic mail: kwyu@phy.cuhk.edu.hk} and
 G. Q. Gu$^{1,3}$}

\address{$^1$Department of Physics, The Chinese University of Hong Kong,
 \\ Shatin, New Territories, Hong Kong \\
 $^2$Institute of Theoretical Physics, The Chinese University of Hong Kong,
 \\ Shatin, New Territories, Hong Kong \\
 $^3$College of Information Science and Technology,
 East China Normal University, \\ Shanghai 200 062, China
}
\maketitle

\begin{abstract}
We have employed the first-principles approach to compute the effective
response of composites of graded spherical particles of arbitrary
conductivity profiles.
We solve the boundary-value problem for the polarizability of the
graded particles and obtain the dipole moment as well as the multipole
moments. We provide a rigorous proof of an {\em ad hoc} approximate
method based on the differential effective multipole moment approximation
(DEMMA) in which the differential effective dipole approximation (DEDA)
is a special case.
The method will be applied to an exactly solvable graded profile.
We show that DEDA and DEMMA are indeed exact for graded spherical particles.
\end{abstract}
\vskip 5mm
\pacs{PACS Number(s): 77.22.-d, 77.84.Lf, 42.79.Ry, 41.20.Cv}

\section{Introduction}

In functionally graded materials (FGM), the materials properties can vary
spatially. These materials have received considerable attention in
various engineering applications \cite{Yamanouchi}.
The variation in the composition yields material and microstructure
gradients, and makes the FGM very different in behavior from the
homogeneous materials and conventional composite materials
\cite{Yamanouchi,Holt}. The great advantage is that one can tailor the
materials properties via the design of the gradients. Over the past few
years, there have been a number of attempts, both theoretical and
experimental, to study the responses of FGM to mechanical, thermal, and
electric loads and for different microstructure in various systems
\cite{Yamanouchi,Holt,Ilschner,fgm1,fgm2,fgm3,fgm4,fgm5}.
In Nature, graded morphogen profiles can exist in a cell layer \cite{fgm1}.
In experiments, graded structure may be produced by using various
approaches, such as a three-dimensional X-ray microscopy technique
\cite{fgm4}, deformation under large sliding loads \cite{fgm5}, and
adsorbate-substrate atomic exchange during growth \cite{fgm3}.
Thus, gradation profiles exist in both natural materials and artificial FGM.
Interestingly, gradation profiles may further be controlled according to
our purpose, such as a specific power-law gradation profile and so on.
It has reported recently that the control of a compatibility factor can
facilitate the engineering of FGM \cite{fgm2}.

There have been various different attempts to treat the composite
materials of homogeneous inclusions \cite{Jackson} as well as multi-shell
inclusions \cite{Gu1,Fuhr,Arnold,Chan}. These established theories for
homogeneous inclusions, however cannot be applied to graded inclusions.
To this end, we have recently developed a first-principles approach for
calculating the effective response of dilute composites of graded
cylindrical inclusions \cite{GuYu} as well as graded spherical particles
\cite{Dong2003}. The electrostatic boundary-value problem of a graded
spherical particle has been solved for some specific graded profiles
to obtain exact analytic results. Along this line, exact analytic
results are available so far for the power-law graded profile
\cite{GuYu,Dong2003}, linear profile \cite{GuYu}, exponential profile
\cite{Martin,Gu} as well as some combination of the above profiles
\cite{Wei}.
For arbitrary graded profiles,
we have developed a differential effective dipole approximation (DEDA)
to estimate the effective response of graded composites of spherical
particles numerically. The DEDA results were shown to be in excellent
agreement with the exact analytic results \cite{Dong2003}.
However, the excellent agreement is difficult to understand because
DEDA method was based on an {\em ad hoc} approximation \cite{Yu02}.

The object of the present investigation is two-fold.
Firstly, we will extend the first-principles approach slightly to deal
with graded particles of arbitrary profiles.
We will solve the boundary-value problem for the polarizability of the
graded particles and obtain the dipole moment as well as the multipole
moments. Secondly, we will provide a rigorous proof of the {\em ad hoc}
approximation from first-principles. To this end, we extend the proof to
multipole polarizability and derive the differential effective multipole
moment approximation. As an illustration of the method, application to
the power-law profile will be made.
Thus, both DEDA and DEMMA are indeed exact for graded spherical particles.

\section{First-principles approach}

In this work, we will focus on a model of a graded conducting particle,
in which the conductivity of the particle varies continuously along the
radius of the spherical particle. We consider the electrostatic
boundary-value problem of a graded spherical medium of radius $a$
subjected to a uniform electric field $E_0$ applied along the $z$-axis.
For conductivity properties, the constitutive relations read
$\vec{J}=\sigma_{i}(r)\vec{E}$ and
$\vec{J}=\sigma_{m}\vec{E}$ respectively in the graded spherical
medium and the host medium, where $\sigma_{i}(r)$ is the conductivity
profile of the graded spherical medium and $\sigma_{m}$ is the
conductivity of the host medium. The Maxwell's equations read
$$
\vec{\nabla} \cdot \vec{J}=0,\ \ \ \ \vec{\nabla} \times
\vec{E}=0.
$$
To this end, $\vec{E}$ can be written as the gradient of a scalar
potential $\Phi$, $\vec{E}=-\vec{\nabla}\Phi$, leading to a
partial differential equation:

\begin{equation}
\vec{\nabla} \cdot [\sigma(r)\vec{\nabla} \Phi] = 0,
\end{equation}
where $\sigma(r)$ is the dimensionless dielectric profile, while
$\sigma(r)=\sigma_i(r)/\sigma_m$ in the inclusion, and
$\sigma(r)=1$ in the host medium.

In spherical coordinates, the electric potential $\Phi$ satisfies
\begin{equation}
\frac{1}{r^2}\frac{\partial}{\partial r} \left(r^2\sigma(r)
\frac{\partial \Phi}{\partial r}\right)
+\frac{1}{r^2\sin\theta}\frac{\partial}{\partial \theta}
\left(\sin\theta \sigma(r) \frac{\partial \Phi}{\partial r}\right)
+\frac{1}{r^2\sin^2\theta}\frac{\partial}{\partial \varphi}
\left(\sigma(r)\frac{\partial \Phi}{\partial \varphi}\right)=0.
\label{comm}
\end{equation}

As the external field is applied along the $z$-axis, $\Phi$ is
independent of the azimuthal angle $\varphi$.
If we write $\Phi = f(r)\Theta(\theta)$ to achieve separation of
variables, we obtain two distinct ordinary differential equations.
For the radial function $f(r)$,
\begin{equation}
\frac{d}{dr}\left(r^2 \sigma(r)
\frac{df(r)}{dr}\right)-l(l+1)\sigma(r)f(r)=0, \label{general}
\end{equation}
where $l$ is an integer, while $\Theta(\theta)$ satisfies the
Legendre equation \cite{Jackson}. Eq.(\ref{general}) is a homogeneous
second-order differential equation; it admits two possible solutions:
$f^{+}_l(r)$ and $f^{-}_l(r)$ being regular at the origin and infinity
respectively. Exact analytic results can be obtained for a power-law
profile \cite{GuYu,Dong2003}, linear profile \cite{GuYu}, and exponential
profile \cite{Martin}. The general solution for the potential in the
spherical medium is thus given by
\begin{equation}
\Phi_{i}(r,\theta) = \sum_{l=0}^{\infty} [A_{l} f^{+}_l(r) + B_{l}
f^{-}_l(r)] P_{l}(\cos\theta).
\end{equation}
In the host medium, the potential is given by
\begin{eqnarray}
\Phi_{m}(r,\theta) = \sum_{l=0}^{\infty} [C_{l} r^l + D_{l}
r^{-(l+1)}] P_{l}(\cos\theta).
\end{eqnarray}
Thus the problem can be solved by matching the boundary conditions
at the spherical surface.

\section{Boundary-value problem}

By virtue of the regularity of the solution at $r=0$, the general
solution inside the particle becomes:
\begin{eqnarray}
\Phi_{i}(r,\theta) = \sum_{l=0}^{\infty} A_{l} f^{+}_l(r)
P_{l}(\cos\theta),\ \ \ r \le a,
\end{eqnarray}
The external potential ($r\ge a$) is:
\begin{eqnarray}
\Phi_{m}(r,\theta) = -E_0 r\cos \theta + \sum_{l=0}^{\infty} D_{l}
r^{-(l+1)} P_{l}(\cos\theta),\ \ \ r \ge a.
\end{eqnarray}
We can rewrite $r\cos \theta$ as $\sum_l \delta_{l1} r^l
P_{l}(\cos\theta)$. Thus $C_l=-E_0\delta_{l1}$ and no multipole moment
will be induced in a uniform applied field.
However, in a nonuniform applied field as in dielectrophoresis of
graded particles, multipole moments will be induced.
Matching boundary conditions at $r=a$,
$$
\Phi_i(a)=\Phi_m(a),\ \ \ \sigma(a)\Phi'_i(a)=\Phi'_m(a),
$$
where prime denotes derivatives with respect to $r$, we obtain a set of
simultaneous linear equations for the coefficients
\begin{eqnarray}
A_l f^+_l(a) &=& -E_0 a^l \delta_{l1} + D_l a^{-(l+1)}, \\
\sigma(a) A_l f^{+\prime}_l(a) &=& -E_0 l a^{l-1} \delta_{l1} -
D_l (l+1) a^{-(l+2)}.
\end{eqnarray}
Solving these equations,
\begin{eqnarray}
A_l &=& -{(2l+1)a^l E_0 \delta_{l1}\over a\sigma(a)
f^{+\prime}_l(a)+(l+1)f^{+}_l(a)} = -{(2l+1)a^l E_0 \delta_{l1}
\over f^+_l(a) [l(F_l+1)+1]},\\
D_l &=& {a^{2l+1} [a\sigma(a) f^{+\prime}_l(a)-lf^{+}_l(a)] E_0
\delta_{l1}\over a\sigma(a) f^{+\prime}_l(a)+(l+1)f^{+}_l(a)} =
{a^{2l+1} l(F_l-1)E_0 \delta_{l1} \over l(F_l+1)+1},
\end{eqnarray}
where
\begin{equation}
F_l={\sigma_i(a)\over \sigma_m}{a f^{+\prime}_l(a) \over l
f^{+}_l(a)}.\label{equivalent}
\end{equation}
When the radial equation is solved for a specified graded profile,
the potential distribution can generally be expressed in terms of
$f^+_l(r)$ and $F_l$. For a uniform applied field, the potential becomes
\begin{eqnarray}
\Phi_{i}(r,\theta) &=& -{3a f^+_1(r) E_0 \over f^+_1(a) (F_1+2)}
\cos\theta,\
\ \ r\le a, \\
\Phi_{m}(r,\theta) &=& -E_0r \cos\theta + {a^3 (F_1-1)E_0 \over
r^2 (F_1+2)} \cos\theta,\ \ \ r\ge a.
\end{eqnarray}
The second term in the potential in Eq.(14) can be interpreted as
the potential due to an induced dipole moment $p=\sigma_m b_1 a^3
E_0$. Thus we identify the dipole factor:
\begin{equation}
b_1={F_1-1\over F_1+2},\ \ \ F_1={\sigma_i(a)\over \sigma_m}{a
f^{+\prime}_1(a) \over f^{+}_1(a)}.
\end{equation}
For a homogeneous (non-graded) particle, the well known result recovers:
$$
b_1={\sigma_i-\sigma_m\over \sigma_i+2\sigma_m}.
$$
Thus, $F_1$ can be interpreted as the equivalent conductivity ratio of
the graded spherical particle.

\section{\bf Differential effective multipole moment approximation}

We should remark that in general $F_l$ is the equivalent conductivity
ratio of the $l$th multipole moment.
Let us extend the definition [Eq.(\ref{equivalent})] slightly for
a graded spherical particle of variable radius $r$:
\begin{equation}
F_l(r)=\sigma(r){r f^{+\prime}_l(r) \over l f^{+}_l(r)}.\label{Fl}
\end{equation}
Thus, the multipole factor reads:
\begin{equation}
b_l(r) = {l(F_l(r)-1)\over l(F_l(r)+1)+1}.
\end{equation}
Physically it means that we construct the graded particle by a
multi-shell procedure \cite{Yu02}: we start out with a graded particle
of radius $r$ and keep on adding conducting shell gradually.
The change in $F_l$ and $b_l$ can be assessed. In this regard, it is
instructive to derive differential equations for $F_l(r)$ and $b_l(r)$.
Let us consider (ignoring the superscript $+$ and subscript $l$ in
$f^+_l(r)$ for simplicity):
$$
{d\over dr}[rF_l(r)] = {1\over l}{d\over
dr}\left[r^2\sigma(r){f^{\prime}(r) \over f(r)}\right] = {1\over
lf(r)}{d\over
dr}[r^2\sigma(r)f^{\prime}(r)]-{r^2\sigma(r)f^{\prime}(r)^2\over l
f(r)^2}.
$$
From Eq.(\ref{general}) and Eq.(\ref{Fl}), we obtain the differential
equation:
\begin{equation}
{d\over dr}[rF_l(r)] = (l+1)\sigma(r) - {lF_l(r)^2\over
\sigma(r)},
\end{equation}
which is just the generalized Tartar formula \cite{Milton}.
From Eq.(17) and Eq.(18), we obtain the DEMMA:
\begin{eqnarray}
{d b_l(r)\over dr} &=& -\frac{1}{(2l+1)r\sigma(r)}
[(b_l(r)+l+b_l(r)l)+(b_l(r)-1)l\sigma(r)] \nonumber \\ & &\times
[(b_l(r)+l+b_l(r)l)-(b_l(r)-1)(l+1)\sigma(r)]. \label{DEMMA}
\end{eqnarray}
When $l=1$, we recover the Tartar formula and DEDA \cite{Yu02}:
\begin{equation}
{d\over dr}[rF_1(r)] = 2\sigma(r) - {F_1(r)^2\over \sigma(r)},
\end{equation}
\begin{eqnarray}
{d b_1(r)\over dr} &=& -\frac{1}{3r\sigma(r)}
[(2b_1(r)+1)+(b_1(r)-1)\sigma(r)] \nonumber \\ & &\times
[(2b_1(r)+1)-2(b_1(r)-1)\sigma(r)]. \label{DEMA}
\end{eqnarray}

Let us consider a graded particle in which the conductivity profile has
a power-law dependence on the radius, $\sigma(r)=cr^k$, with $k \ge 0$
where $0 < r \le a$. Then the radial equation becomes
\begin{equation}
\frac{d^2
f}{dr^{2}}+\frac{k+2}{r}\frac{df}{dr}-\frac{l(l+1)}{r^2}f=0.
\label{Re}
\end{equation}
As Eq.(\ref{Re}) is a homogeneous equation, it admits a power-law
solution \cite{Dong2003},
\begin{equation} f(r)=r^{s}.
\end{equation}
\label{f(r)} Substituting it into Eq.(\ref{Re}), we obtain the
equation $s^{2}+s(k+1)-l(l+1)=0$ and the solution is
\begin{equation}
s^{k}_{\pm}(l)=\frac{1}{2}\left[-(k+1)\pm\sqrt{(k+1)^{2}+4l(l+1)}
\right]. \label{sp}
\end{equation}

There are two possible solutions:
$$
f^{+}_l(r)=r^{s^{k}_{+}(l)},\ \ \ \ f^{-}_l(r)=r^{s^{k}_{-}(l)}.
$$
Thus the $l$-th order equivalent conductivity becomes:
\begin{equation}
F_l = {\sigma_i(a)\over \sigma_m}{af^{+\prime}_l(a) \over
l f^{+}_l(a)} = \frac{s^k_+(l)}{l}ca^k. \label{power}
\end{equation}
When $k \to 0$, $s^k_{+}(l) \to l$, $F_l \to c$, the result for
a homogeneous sphere recovers.

\section{Discussion and conclusion}

Here a few comments are in order.
We have employed the first-principles approach to compute the multipole
polarizability of graded spherical particles of arbitrary conductivity
profiles and provided a rigorous proof of the differential effective
dipole approximation.

We are now in a position to propose some applications of the present
method. We may attempt the similar calculation of the multipole response
of a graded matallic sphere in the nonuniform field of an oscillating
point dipole at optical frequency. The graded Drude dielectric function
will be adopted \cite{APL}.
The similar approach may also be extended to anisotropic medium with
different radial and tangential conductivities \cite{Dong2004}.
Similar work can be extended to ac electrokinetics of graded cells
\cite{Huang2004}. We can also study the interparticle force between
graded particles \cite{Yu}.

In summary, we have solved the boundary-value problem for the
polarizability of the graded particles and obtained the dipole moment as
well as the multipole moments. We provided a rigorous proof of the
differential effective multipole moment approximation. We showed that
DEDA and DEMMA are indeed exact for graded spherical particles.
Note that an exact solution is very few in composite research and to have
one yields much insight. Such solutions should be useful as benchmarks.
Finally, we should remark that the exact derivation of DEDA and DEMMA is
for graded spherical particles only. For graded nonspherical particles,
these {\em ad hoc} approaches may only be approximate.

\section*{Acknowledgments}

This work was supported by the Research Grants Council Earmarked
Grant of the Hong Kong SAR Government, under project number CUHK
403004. G.Q.G. acknowledges support by Natiaonal Sciene Foundation
of China, under Grant No. 10374026. K.W.Y. acknowledges useful
discussion and fruitful collaboration with L. Dong, J. P. Huang,
Joseph Kwok, J. J. Xiao and C. T. Yam.

\end{document}